\documentclass[conference]{IEEEtran}
\IEEEoverridecommandlockouts

\usepackage[latin1]{inputenc}
\usepackage[cmex10]{amsmath}
\usepackage{amsfonts}
\usepackage{amssymb}
\usepackage{graphicx}
\usepackage{verbatim}
\usepackage{array}
\usepackage{booktabs}
\usepackage{multicol}
\usepackage{multirow}
\usepackage{dcolumn}
\usepackage{color}
\usepackage[noadjust]{cite}
\usepackage{url}
\usepackage{balance}
\usepackage[usenames,dvipsnames]{xcolor}
\usepackage{algorithm}
\usepackage[noend]{algpseudocode}
\usepackage[flushleft]{threeparttable}
\usepackage{siunitx}
\usepackage{mathtools}

\DeclareGraphicsExtensions{.eps}

\makeatletter
\def\BState{\State\hskip-\ALG@thistlm}
\makeatother

\begin{document}

\title{Peer-group Behaviour Analytics of Windows Authentications Events Using Hierarchical Bayesian Modelling\\
\thanks{This research was funded by Securonix Threat Labs, Securonix Inc. Austin, Texas. Corresponding author: hhhelfer@securonix.com. $^{*}$This work was completed as part of an internship at Securonix. $^{\dag}$Equal contribution.}
}

\author{\IEEEauthorblockN{Iwona~Hawryluk$^{a, b, *, \dag}$, Henrique~Hoeltgebaum$^{b,\dag}$, Cole~Sodja$^{b}$, Tyler~Lalicker$^{b}$ and Joshua~Neil$^{b}$}
\IEEEauthorblockA{$^{a}$\textit{Department of Infectious Disease Epidemiology, School of Public Health, Imperial College London, UK} \\
$^{b}$\textit{Securonix Threat Labs, Securonix Inc., USA}
}}

\maketitle
\begin{abstract}
 
Cyber-security analysts face an increasingly large number of alerts received on any given day. This is mainly due to the low precision of many existing methods to detect threats, producing a substantial number of false positives. Usually, several signature-based and statistical anomaly detectors are implemented within a computer network to detect threats. Recent efforts in User and Entity Behaviour Analytics modelling shed a light on how to reduce the burden on Security Operations Centre analysts through a better understanding of peer-group behaviour. Statistically, the challenge consists of accurately grouping users with similar behaviour, and then identifying those who deviate from their peers. This work proposes a new approach for peer-group behaviour modelling of Windows authentication events, using principles from hierarchical Bayesian models. This is a two-stage approach where in the first stage, peer-groups are formed based on a data-driven method, given the user's individual authentication pattern. In the second stage, the counts of users authenticating to different entities are aggregated by an hour and modelled by a Poisson distribution, taking into account seasonality components and hierarchical principles. Finally, we compare grouping users based on their human resources records against the data-driven methods and provide empirical evidence about alert reduction on a real-world authentication data set from a large enterprise network.

\end{abstract}


\section{Introduction}\label{Intro}

\IEEEPARstart{T}{he} increasingly complex attack methods adopted by malicious entities to evade existing defences in cyber environments are a growing concern for society. A recent survey points towards losses of around \$600 billion per annum \cite{lewis2018economic}. This is deeply troubling since large-scale enterprises heavily depend on computers and communication networks in order to function \cite{kitchin1998cyperspace}. The security of such networks is usually implemented and periodically updated by domain experts. However, security is becoming a significant concern due to the highly sophisticated threat technology adopted by malicious actors. This means that security specialists are forced to act in a reactive manner \cite{dhir2021prospective}.

This is very restrictive, as the majority of operational security systems rely on signature-based or rule-based methods, which look for events and behaviours of known attack form \cite{turcotte2017unified}. These methods are unable to handle the so-called "zero-day" attacks, i.e., activities that have not previously been reported \cite{patcha2007overview}. In addition, an increasing number of cyber threats provide evidence that these pre-defined, static methods are not sufficient and lag behind the attackers' sophistication. This has boosted the research area which focuses on using machine learning and statistical models to perform anomaly detection in computer network data. Those methods aim to estimate a "normal" behavioural profile and assess whether an entity (for instance user, host or computer) is deviating with respect to its peers \cite{heard2016dynamic}. In industry, such techniques are commonly referred to as User and Entity Behaviour Analytics (UEBA). This peer-group analysis is vital in terms of false alarm reduction. In cyber-security applications, peer-groups are usually determined based on human resource (HR)
records, according to organisational structure. This is restrictive, since users having the same job title could perform different activities related to their group, not necessarily triggering an alert, and can also lead to false positives when peer groups are ill-defined.

Regarding statistical models in cyber-security, authentication data is a rich source of information to perform anomaly detection. Several works have explored the idea of identifying anomalous connections between entities using graph-based methods (e.g. \cite{neil2013towards, kent2015authentication, turcotte2014detecting,turcotte2016poisson, passino2022graph}). Whereas other approaches focused on modelling authentications counts logs, adopting methods that are able to handle streaming data (\cite{lambert2006adaptive, Hoeltgebaum2021}).

Due to the need for these algorithms to define baseline behaviour, most of the anomaly detection results are strongly dependent on the clustering method used to aggregate users of similar logging patterns \cite{turcotte2016poisson}. However, there simply does not exist a panacea for clustering methods with regard to authentication modelling. Several methods have been explored, each exposing its own weaknesses and strengths. Recently, in \cite{metelli2019bayesian} the authors presented a robust clustering method based on a relevant set of covariates. This work was based on a former Bayesian Cox model from the same authors \cite{metelli2016model} to perform prediction of new edges in the computer network and anomaly detection. Initial cluster configurations were based on the spectral bi-cluster from \cite{dhillon2001co}, which is also explored in our work.

Bayesian models are becoming increasingly popular among cyber-security applications, especially when modelling user behaviour within a computer network (see \cite{perusquia2022bayesian} for a complete review). The main advantage of using Bayesian methods lies in their capability to quantify the uncertainty of the predictions and include prior knowledge, allowing the practitioners to make more informed decisions even when little data is available. Additionally, it allows to perform Bayesian hierarchical modelling, which is based on the principle of borrowing strength via hierarchical estimation \cite{congdon2019bayesian}. The idea is to perform inference for a collection of exchangeable units using Bayesian Hierarchical methods (henceforth referred to as BH). There is a clear demand for the development and use of BH in applied problems from various areas, from drug development \cite{gupta2012use} to sports analytics \cite{kruschke2015bayesian} and many others. Recently, this framework had a great impact throughout the COVID-19 pandemic, where advances in BH modelling allowed the estimation of the reproduction number of the virus even at the early stages when the availability of the data was largely limited \cite{flaxman2020estimating}.

In this work, we propose a framework for modelling authentication data and identifying anomalous spikes in the number of unique target entities that users authenticate to. This is motivated by a particular stage of the \emph{kill chain} \cite{hutchins2011intelligence} called ``reconnaissance'', in which a malicious actor tries to connect to several entities within a network, trying to identify vulnerable ports or confidential data location \cite{recon}. Specifically, the contributions of our paper are the following: (i) Five methods to cluster users based on their behaviour are evaluated, (ii) Six Bayesian approaches to model authentications are proposed, including explicit effects of weekly and hourly seasonality and hierarchical prior for an authentication method, (iii) A full Bayesian framework to conduct anomaly detection, based on the highest posterior density interval is proposed, taking into account the full posterior distribution rather than only the expected values of the parameters. (iv) The analysis is conducted on a real-world data set of authentication logs, provided by a non-disclosure source for research purposes.

The paper is structured as follows: Section~\ref{Data} gives a high-level description of windows authentication events and conducts a brief exploratory data analysis (EDA) on the data we then model. Section~\ref{cluster} details the peer-group clustering approaches. Section~\ref{hierarc} presents six Bayesian time series models, along with the diagnostics and convergence details. Section~\ref{anom} provides details on the anomaly detection framework used here, while the results are given in Section~\ref{case_study}. Finally, some conclusions and future work are described in Section ~\ref{concl}.


\section{Data: Windows Authentication Events}\label{Data}

As the main foundation for our analysis, an authentication data set from a specific enterprise is used. Due to non-disclosure agreements, any source of identifying information to the data owner is omitted. The events logs used here come from two sources, (i) centralised authentication servers (Microsoft Active Directory servers) and (ii) individual networked computers running the Microsoft Windows operating systems (desktops and servers). Here we only consider logs from (ii) and, analogous to \cite{kent2015authentication}, only successful authentication events originating from Kerberos and Microsoft's NTLM are considered. In addition, only human-account activities were considered, so the user names ending with ``\$'' or matching \emph{LOCAL}, \emph{SYSTEM}, \emph{ANONYMOUS}, and \emph{ADMINISTRATOR} were removed.

\begin{figure*}[htbp]
\centering
\includegraphics[width=1.5\columnwidth]{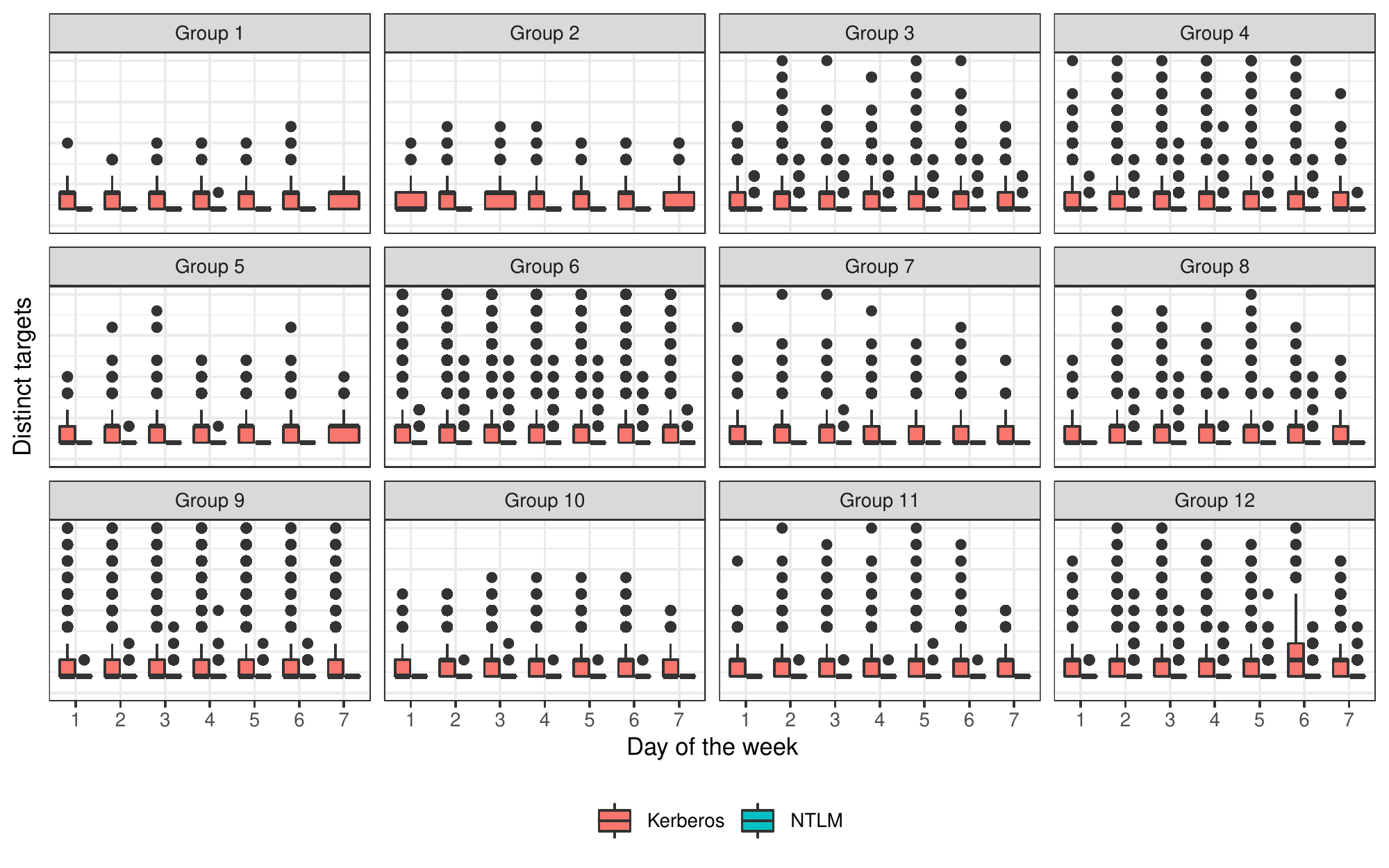}
\includegraphics[width=2.0\columnwidth]{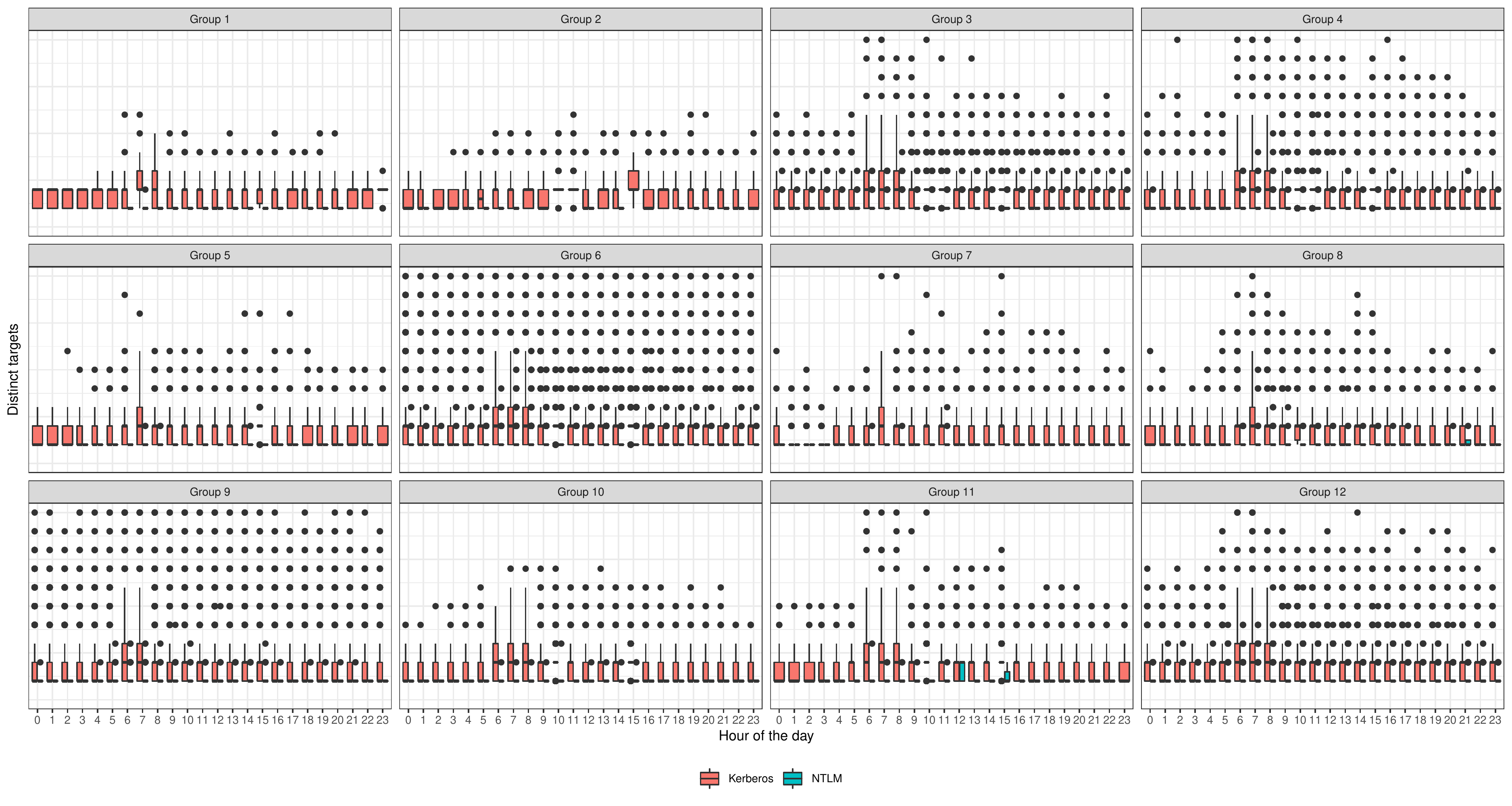}
\caption{Weekly (top) and hourly (bottom) behaviour patterns of users in an enterprise. In the top plot, day 1 denotes Sundays, while in the bottom plot, hour 0 denotes midnight. For illustration purposes, the y-axes were capped and the values were omitted.}
\label{season_features}
\end{figure*}

The data set considered here spans a recent period of 27 days. To avoid issues about having users from different time zones, we consider only employees from a single country. Further, the data is aggregated at an hourly level, counting the number of distinct entities that user $u$ authenticated to at timestamp $t$. Throughout this paper, we refer to this unique number of target entities an user authenticates to simply as \emph{authentications}. Fig.~\ref{season_features} shows an example of weekly and hourly patterns in users' authentication behaviour in the considered data set. The users are aggregated based on their HR divisions which are encoded into enumerated groups to preserve privacy. Fig.~\ref{season_features} reveals that certain groups clearly have daily patterns, i.e. they are mostly authenticating during working hours, while for others this distinction is less clear. Similarly, for the weekly patterns, there are groups which authenticate mostly on working days, while for others this is not so clear.

For fitting and evaluation purposes, the data is split into a train set (first 20 days) and a test set (last 7 days). The proposed models are fitted to the training sample while anomaly detection is performed by comparing predictions from those models to the observed values within a testing sample. Users who have observations in the test set but not in the training set are removed. In addition, users with less than 10 observations in the training set are also removed from the analysis. Applying those criteria we end up with 7504 users and 400 timestamps in the training set and 168 timestamps in the test set. While this is a relatively small sample in cyber-security, the point of this paper is to exemplify the power of BH in small sample settings. Such settings exist when models are first deployed in an enterprise, or when new users appear, which happens very often, and is referred to as the cold start problem among practitioners.

\section{Clustering users by authentication profile}\label{cluster}

As argued earlier, there is no optimal data-driven method to create peer-groups when dealing with authentication data sets. The definition of groups is vital in order to borrow strength from the right peers, whereas a poor grouping method has the potential to harm the inference regarding anomaly detection within the group. In this section, five clustering approaches are explored to analyse user-authentication behaviour. The first approach is based on HR records and the remaining four are purely data-driven. Taking the training sample, assume both sets of users and destination targets are numbered, i.e, $U=\{1,...,|U|\}$ and $C=\{1,...,|C|\}$. Denote the adjacency matrix $A\in \{0,1\} ^ {|U| \times |C|}$ as the connectivity pattern from the user to the destination targets, where after observing $n$ edges $(u_1,c_1), ..., (u_n,c_n)$ each entry is defined as $A_{u,c} = \sum_{i=1}^{n} \mathbf{1} \{(u_i,c_i)\}$. The data-driven clustering algorithms are defined as the following.

\begin{itemize}
    \item TS: Automatically fitting univariate ARIMA (autoregressive integrated moving average) models to each user count time series. The order of each model is selected based on the one minimising AIC (Akaike Information Criterion). The residuals of each model are then clustered based on a Gaussian Mixture Model (GMM);
    \item k-means: Calculate a Singular Value Decomposition (SVD) of $A$ and then apply a k-means clustering algorithm on the left singular vectors which correspond to the users. In order to determine the optimal number of clusters, the elbow method is used such that the total intra-cluster variation is minimised \cite{elbow};
    \item GMM: Calculate a SVD of $A$ and then apply a Gaussian Mixture Model algorithm on the left singular vectors which correspond to the users;
    \item spectral Bi-cluster: Denote $D_U$ and $D_C$ as diagonal matrices with row and column sums of A respectively. These two former quantities provide an idea of the indegree of targets and the outdegree of users. This approach is based on the spectral bi-clustering algorithm from \cite{dhillon2001co} which calculates a truncated-singular value decomposition on $D_U^{-1/2} A D_C^{-1/2}$.
\end{itemize}

\section{Hierarchical modelling}\label{hierarc}
In this section, we present six Bayesian models used for modelling the authentication counts. Recall, that in this paper we refer to authentications as the number of unique target entities a user authenticates to in a given hour, denoted as $y$. Each model tries to capture a dependency structure observed in the data while conducting an EDA (Fig.~\ref{season_features}). The natural candidate to model count data with overdispersion is the Poisson-Gamma model. Nevertheless, while evaluating the authentication counts for each user, overdispersion was not present in this data set, i.e., $V(y)/E(y)$ did not exceed 1. For this reason, only the Poisson model is considered here. It is also worth mentioning, that each user's time series has a different length, as only events different from zero were recorded, removing zero-inflation.

In the first two models, a complete pooling structure is assumed, i.e., no grouping structure is adopted. Denote $y$ as the authentication counts aggregated per hour, taking each user's time series as an \emph{i.i.d.} event, model 1 is defined as
\begin{eqnarray}
y &\sim& \text{Poisson}(\lambda)\nonumber\\
\text{log}(\lambda) & \sim & N(0,5)\,.\label{eq: model1}
\end{eqnarray}
\noindent
Nevertheless, as illustrated in Fig.~\ref{season_features}, the seasonality component has a strong effect throughout working hours and days. With this feature in mind, model 2 treats the counts of each user per days $d=1,...,7$ and hours $h=1,...,24$ as \emph{i.i.d.}, rendering $\Lambda \in \mathbb{R}^{24 \times 7}$ parameters and estimated as
\begin{eqnarray}
y_{h,d} &\sim& \text{Poisson}(\lambda_{h,d})\nonumber\\
\text{log}(\lambda_{h,d}) & \sim & N(0,5)\,,\label{eq: model2}
\end{eqnarray}
\noindent with $\lambda_{h,d}$ corresponding to the element of row $h$, column $d$ in $\Lambda$.

Having in mind the UEBA component of this work, the last four models are based on analysing peer-group behaviour. In model 3 a grouping structure is considered, based on HR divisions or one of the clustering methods described in Section~\ref{cluster}.
Counts within each group are treated as \emph{i.i.d} events. Denoting $g \in G$ as the group index and set respectively, model 3 estimates $\Lambda \in \mathbb{R}^{|G|}$ as
\begin{eqnarray}
y_{g} &\sim& \text{Poisson}(\lambda_g)\nonumber\\
\text{log}(\lambda_g) & \sim & N(0,5).\,\label{eq: model3}
\end{eqnarray}

Analogously to model 2, model 4 explicitly captures the seasonality dynamics per group. Define the tensor $\Lambda \in \mathbb{R}^{24 \times 7 \times |G|}$, treating the observations for each group $g \in G$, at hour $h$ and day $d$ as \emph{i.i.d}, model 4 is defined as
\begin{eqnarray}
y_{h,d,g} &\sim& \text{Poisson}(\lambda_{h,d,g})\nonumber\\
\text{log}(\lambda_{h,d,g}) & \sim & N(0,5).\,\label{eq: model4}
\end{eqnarray}

After a closer inspection of Fig.~\ref{season_features}, it is possible to notice a different behaviour in the range of values assumed by the authentication method. To capture this dynamic, model 5 adds an additional coefficient matrix $\Psi \in \mathbb{R}^{2 \times |G|}$ which corresponds to the authentication method Kerberos or NLTM, $m=\{1,2\}$. The tensor defined in model 4 is adopted and model 5 can be written as
\begin{eqnarray}
y_{h,d,g,m} &\sim& \text{Poisson}(\lambda_{h,d,g} \times \psi_{m,g})\nonumber\\
\text{log}(\lambda_{h,d,g}) & \sim & N(0,5)\nonumber\\
\text{log}(\psi_{m,g}) & \sim & N(0,5)\,\label{eq: model5}
\end{eqnarray}
\noindent where $\psi_{m,g}$ corresponds to the elements of the matrix $\Psi$, such that $\psi_{1,g}$ denotes the effect of Kerberos authentication for group $g$, whereas $\psi_{2,g}$ denotes the effect of NTLM authentication for the same group.

Moreover, if fixed effects models were considered, not splitting the effect of method (Kerberos or NTLM), the average behaviour is very different, which is possible to infer trough a visual inspection. Hierarchical random effects would pool this information across units in order to obtain the most accurate estimate for each method. This idea leads to Model 6, which adds a hierarchical structure to the method coefficient as
\begin{eqnarray}
y_{h,d,g,m} &\sim& \text{Poisson}(\lambda_{h,d,g} \times \psi_{m,g})\nonumber\\
\text{log}(\lambda_{h,d,g}) & \sim & N(0,5)\nonumber\\
\text{log}(\psi_{m,g}) & \sim & N(\mu_{m},5)\nonumber\\
\mu_{m} & \sim & N(0,5)\,\label{eq: model6}
\end{eqnarray}

Posterior samples of the parameters in the models were generated using Stochastic Variational Inference (SVI) with NumPyro (v. 0.10.0) \cite{numpyro, pyro}. Usage of Hamiltonian Monte Carlo (HMC) was also investigated, however, due to the large size of the dataset (over 1M rows) SVI was deemed more appropriate due to time consideration. For example, fitting model 4 took 2.5 minutes with SVI compared to 51 minutes with HMC on 6 AWS (Amazon Web Services) servers with 16 cores and 128 GB of RAM. For the SVI optimisation, 5000 steps with Adam optimiser \cite{adam} were run with a fixed learning rate of $0.01$. The convergence of each model fit was evaluated by ensuring that the convergence statistic $\hat{R}$ was less than $1.01$ for each parameter \cite{rhat}.


\section{Anomaly detection}\label{anom}

Typically, anomaly detection using similar models would consist of calculating the probabilities of observing value $y$ assuming a Poisson distribution with rate parameter $\lambda$. These probabilities take into account the expected value of the posterior distribution $E[\lambda|y]$. Such an approach does not take into consideration the advantages of a fully Bayesian framework, as it is using only point estimates rather than the uncertainty from the whole posterior distribution.

Instead, once the model parameters are estimated, the posterior predictive distribution can be evaluated. With this distribution, the highest posterior density interval (HPDI), with significance level $\alpha$ \cite{hpdi}, is calculated. HPDI is a region that contains $100(1-\alpha)\%$ of the posterior distribution's mass and can be interpreted as the credible interval (in fact, if the posterior is symmetric and unimodal credible interval and HPDI are the same). If a value is observed outside this interval, it can be flagged as an anomaly. Here we are assuming that observed spikes in the authentication counts are associated with malicious events, so only HPDI's upper limit is considered.

Formally, define $\mathbf{Y}=\{y_{1}, .., y_{T}\}$ as the authentication counts in the training set and $\theta_j$ as the estimated parameter vector for model $j = \{1,...,6\}$. Value $y$ is flagged as an anomaly if $y > HPDI_{\alpha}^{\text{UP}}f(\theta_j|\mathbf{Y}))$. Here, $\text{HPDI}_{\alpha}^{\text{UP}}$ denotes the upper limit of the $100(1-\alpha)\%$ HPDI and $f(\theta_j|Y_u)$ denotes the posterior predictive distribution given the parameter's $\theta_j$ distribution. Note that the specific form of $f$, the shape of the parameter vector $\theta_j$ and the training set structure depends on which of the six models was applied. Moreover, this approach for anomaly detection is easily extendable to other Bayesian models and datasets.

\begin{figure}[t!]
\centering
\includegraphics[width=0.95\columnwidth]{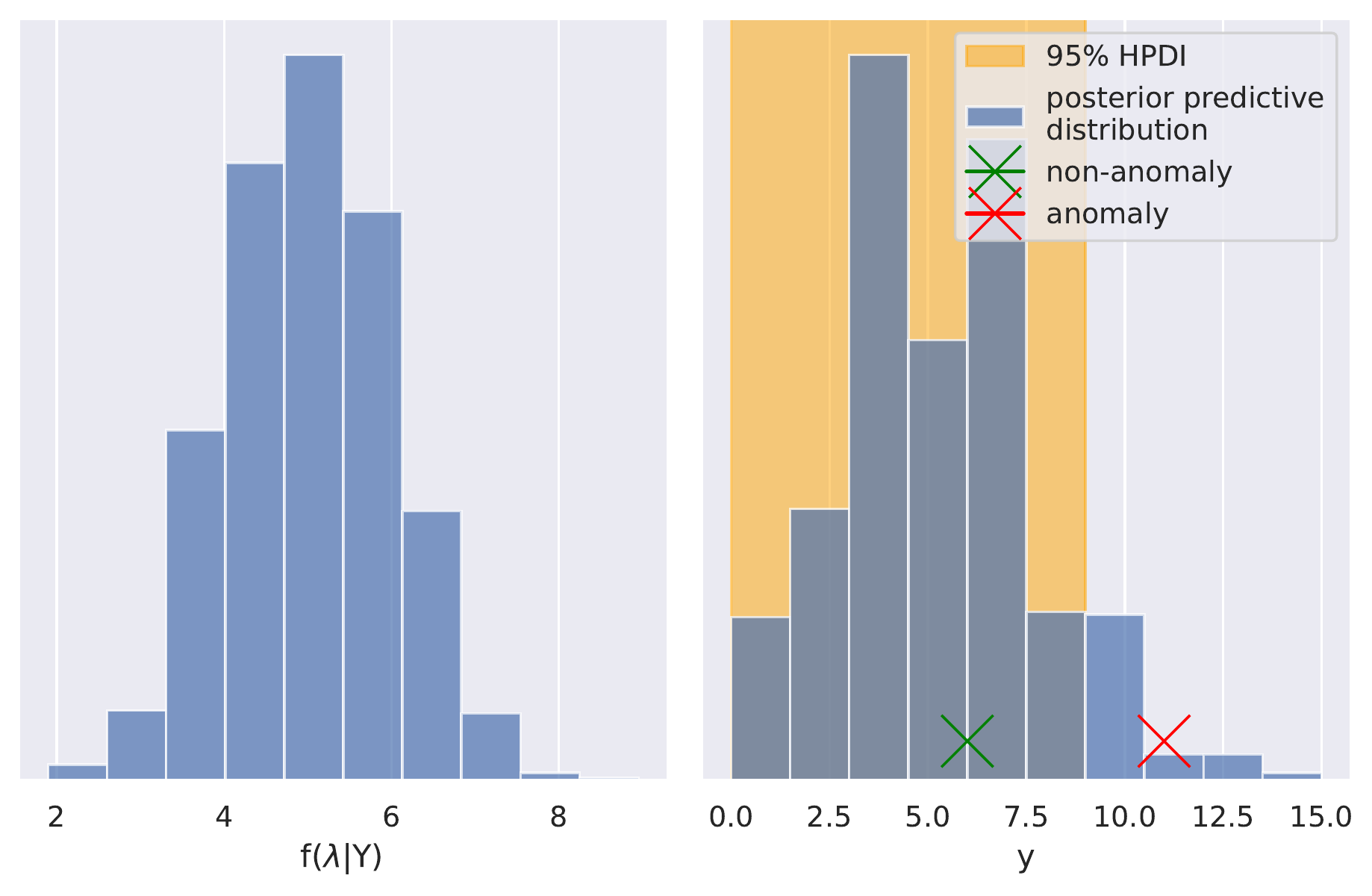}
\caption{Example of anomaly detection using HPDI. Left: posterior density of random variable $\lambda$. Right: posterior predictive distribution, evaluated as $\text{Poisson}(\lambda)$ and 95\% HPDI. Two values are shown as examples of non-anomaly (value 6) and anomaly (value 11).}
\label{example_anomaly}
\end{figure}

An example of this approach is illustrated in Fig.~\ref{example_anomaly}, where we generated a posterior predictive distribution for a distribution of a parameter $\lambda$ and then calculated 95\% HPDI, as shown with the shaded area. Next, two values are given: one non-anomalous, shown as a green cross and within the HPDI, and one anomalous, with a value higher than the upper limit of HPDI and shown with a red cross.

\section{Case study}\label{case_study}

The six competing models along with the peer-group modelling approaches are applied to the enterprise authentication logs data. In practice, Security Operation Centre (SOC) analysts receive a high number of alerts, of which the majority end up being a false detection. In addition, as with most post-breach detection scenarios, the data is not labelled, that is we do not have any information regarding if a true attack took place or not. With such claims in mind, here the approaches will be compared in terms of alert rate, i.e., the proportion of alerts that each model produces in the test set.

Results of the anomaly detection analysis are given in Table~\ref{detect_perf}. This table shows the percentage of all observations in the test set which are labelled as an anomaly by the methodology described in Section~\ref{anom}, using the 99\% HPDI. The total number of observations evaluated in the test set are 439,689. Additionally, the number of clusters provided by each clustering algorithm is also provided in the last row. Note that results for models 1 and 2 are the same, regardless of the grouping, since the models are completely pooled (see Section~\ref{hierarc}). It is worth recalling that the motivation of the paper is to reduce the burden on SOC analysts. This means that we are favouring approaches that render the smallest alert rates. With that in mind, we favour the models producing the lowest alert rates, which are achieved by models 5 and 6 with the GMM grouping method. Also, note that these alert rates are even smaller than for the default grouping obtained through the HR records. This is expected as employees in the same division do not necessarily have the same connectivity pattern. Also, it is not a rare event that employees change their job title within the company or are allocated to other divisions due to budget constraints but keep working in a completely different one. Moreover, the benefit of using the data-driven approaches to users grouping allows to perform similar analyses even when the HR records are not available, or are of a poor quality, which has large benefits for deployment and maintenance of detection technology.

\begin{table}[t]
\caption{Alert rates for different models and grouping methods evaluated on the test set. Any observation above of the 99\% HPDI is considered to be an anomaly. Table rows make reference to the different models while columns refer to the grouping strategies.\label{tab: results_alerts_99}}
\begin{tabular}{llllll}
\toprule
{} & HR & TS & k-means & GMM & Bi-cluster \\
\midrule
M1  &         1.10\% &         1.10\% &             1.10\% &         1.10\% &                 1.10\% \\
M2  &         0.70\% &         0.70\% &             0.70\% &         0.70\% &                 0.70\% \\
M3  &         0.70\% &         0.70\% &            0.71\% &        0.69\% &                0.61\% \\
M4  &        0.63\% &        0.68\% &            0.66\% &        0.63\% &                0.63\% \\
M5  &         0.60\% &        0.63\% &             0.60\% &        0.57\% &                 0.60\% \\
M6  &        0.61\% &        0.62\% &            0.59\% &        0.57\% &                0.61\% \\
\hline
N groups &           12 &            3 &               15 &            8 &                    3 \\
\bottomrule
\end{tabular}
\label{detect_perf}
\end{table}

The widely applicable information criterion (WAIC) is a popular method for Bayesian model selection and is applied here for each of the proposed models. WAIC allows us to estimate out-of-sample predictive accuracy, using the log-pointwise posterior predictive density \cite{waic, loo-cv}. Results regarding WAIC are shown in Table~\ref{waic_table}, where a higher WAIC value indicates better out-of-sample performance. The results suggest that the hierarchical structure improved model fit, regardless of the clustering method. Considering the clustering structure, the GMM performed better, followed by the bi-spectral clustering, which advocates in favour of grouping the users' connection patterns through the adjacency matrix.

\begin{table}[t]
\caption{WAIC for each proposed combination of clustering methods and models. All values are on a log scale and higher is better. \label{tab: results_waic}}
\begin{tabular}{lrrrrr}
\toprule
{} &  HR &  TS &  k-means &  GMM &  Bi-cluster \\
\midrule
M1 &      -1452376 &      -1452376 &          -1452376 &      -1452376 &              -1452376 \\
M2 &      -1441106 &      -1441106 &          -1441106 &      -1441106 &              -1441106 \\
M3 &      -1449765 &      -1449973 &          -1449231 &      -1444705 &              -1442660 \\
M4 &      -1438803 &      -1438946 &          -1438793 &      -1433673 &              -1432262 \\
M5 &      -1427199 &      -1427456 &          -1425775 &      -1419099 &              -1421899 \\
M6 &      -1427017 &      -1427274 &          -1425461 &      -1418925 &              -1421392 \\
\bottomrule
\end{tabular}
\label{waic_table}
\end{table}

In addition, model diagnostics were evaluated through quantile residuals analysis \cite{dunn1996randomized}. Under correct model specification, these residuals should be normally distributed and show no temporal dependence. These can be checked, for example, using the standard Jarque--Bera test for normality and the Ljung--Box test for the absence of serial correlation. Even though a few rejections for both tests happened in some models, on average the models provided evidence that they were capable of filtering the dependency structure in these count time-series. To improve model diagnostics, dummy variables on abnormal values should be introduced on a user basis or extra covariates should be considered. We leave this as future work, where ideas from shrinkage could be adopted.

\section{Conclusions}\label{concl}
A new approach to performing anomaly detection in counts of unique target entities was proposed in this work. The approach is based on Bayesian modelling, incorporating time-effects and hierarchical principles in addition to peer-group behaviour modelling. The results of our proposed method on a real-world dataset highlighted the benefits of our approach in terms of alert rates. Additionally, the grouping methods based on data-driven approaches showed good performance without relying on availability of the HR records. Unfortunately, no labels were available to evaluate a true detection. In addition, nearly one month of data was available, hence any multi-month cycle effect was not possible to model. For future work we believe that shrinkage approaches could be applied for variable selection, improving the model diagnostics. In addition, as the alert rates are strongly dependent on the clustering methods, shrinkage approaches could also be applied to propose a better-suited hierarchy.

\vspace{-0.2cm}
\bibliographystyle{IEEEtran}
\bibliography{references}

\end{document}